\documentclass[fleqn,11pt]{article}
\usepackage{graphicx,amssymb}

\newcommand{\AmS}{{\protect\the\textfont2
  A\kern-.1667em\lower.5ex\hbox{M}\kern-.125emS}}
\textheight by \topskip \textwidth 455pt \hfuzz=15pt
  \normalsize
\begin{document}
\centerline{\bf THERMAL NEUTRONS IN EAS: A NEW DIMENSION IN EAS
STUDY\footnote{Talk given at the ISVHECRI'2006, Weihai, China}}
\vspace*{0.2truein} \centerline{\footnotesize YURI V.
STENKIN\footnote{e-mail: stenkin@sci.lebedev.ru}}
\baselineskip=12pt \centerline{\footnotesize\it  Institute for
Nuclear research of Russian Academy of Sciences}
\baselineskip=10pt \centerline{\footnotesize\it 60th October
anniv. prospect, 7a, Moscow 117312, Russia}
\begin{abstract}
A new method to study Extensive Air Shower (EAS) hadronic
component is proposed. It is shown that addition of specific
detectors for thermal neutron detection to a standard array for
EAS study can significantly improve its performance. Results of
CORSIKA based Monte Carlo simulations as well as preliminary
experimental data are presented. A proposal of novel type of EAS
array is given.
\vspace{1pc}
\end{abstract}

\section{Introduction}

The study of cosmic ray neutron component started in 30-s of last
century. A summary of the early experimental data and their
interpretation can be found in $\cite{bet}$. Later, in 40-s, in
parallel with understanding of the EAS hadronic structure,
measurements of neutron component in EAS have began %
$\cite{ton,coc1,coc2}$. All the data obtained in these early works
were correctly understood and interpreted. Later, when the neutron
monitors $\cite{hat}$ were constructed and spread widely, people
tried to use them in conjunction with EAS arrays to study hadronic
component of secondary cosmic rays (see for example %
$\cite{koz,nie,act,kal}$). Many ``anomalies'' were observed in
these experiments: in hadron spectrum $\cite{nie}$, in lateral
distribution $\cite{act}$, etc. Explanations of these
``anomalies'' can be found in an {\it a priori} assumption that
they recorded a single hadron. But, with primary energy rising,
there will be a moment when the number of hadrons entering the
monitor becomes bigger then 1. This is a new class of events,
which we called as {\it hadron group} $\cite{ste1}$. Starting from
this moment, all secondary processes including evaporation neutron
production, depend mostly on the hadrons number reached the
detector instead of their energy rising very slowly. This results
in sharp changes in many observables: locally produced neutrons
number distribution becomes flatter, their lateral distribution
becomes difficult for interpretation while it remains constant for
each center of generation (for each interacting hadron).

The idea to use neutrons moving with sub-luminal velocity at a
long distances from EAS core for the estimation of hadronic
component energy, has been proposed by J. Linsley $\cite{lin}$.

The idea to use thermal neutrons as a key to select muon hadronic
interactions underground has been proposed and realized by
G.T.Zatsepin and O.G.Ryazhskaya $\cite{zat,bez}$

\section{A prototype of the MultiCom array}

A novel type of an array for EAS study (MultiCom) prosed by us in
2001 $\cite{ddd,svg}$, has been realized in 2005 near the existing
Baksan Carpet-2 EAS array as a prototype, consisting of one
working module of $5 \times 5 m^{2}$. 4 thick liquid scintillator
detectors $(70\times 70\times 60 cm^{3})$ in the corners were used
for triggering with a threshold of 106 MeV in each detector. The
trigger counting rate is equal to 3.3 $min^{-1}$. Additional
requirement for event to be stored (software trigger) is energy
deposit in the central ZnS detector equivalent to 1/2 of the most
probable neutron pulse height (or $\sim$8 relativistic particles).
Corresponding energy threshold for such trigger conditions was
calculated to be $\sim7$ TeV for proton originated EAS, $\sim30$
TeV for He EAS and $\sim$300 TeV for Fe EAS. In the center of
module there is situated unshielded thermal neutron detector of
$0.7 m^{2}$ at 2.5 m above ground level (fig. 1). We used a thin
layer of a mixture of old inorganic scintillator ZnS(Ag) with LiF
enriched with $^{6}Li$ up to $90\%$.
\begin{figure}[!htb]
\hspace{-0.75pc} \begin{center}
\includegraphics[width=9cm]{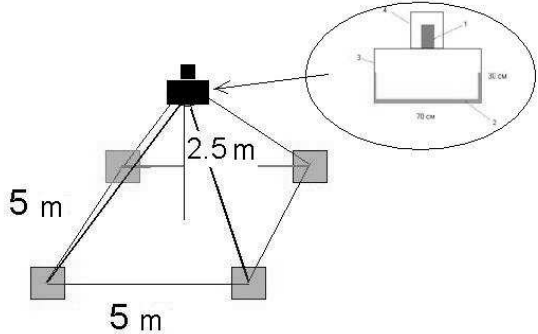}
\caption{MultiCom prototype set-up.}
\end{center}
\end{figure}
%
 Thermal neutrons are recording due to
$^{6}Li(n,\alpha)^{3}H + 4.78$ MeV reaction. ZnS scintillator is
the best scintillator for heavy particle detection and produces
$\sim160,000$ light photons per one captured neutron. That means
one could make a large detector viewed by a single PMT and have
enough light. In our case we have $\sim$50 photo-electrons from
PMT photo-cathode.

The efficiency of thermal neutron detection was found to be
$20\%$. Pulse duration (the fastest component) is equal to $\sim
40$ ns. Due to very thin scintillator layer ($~30$ $mg/cm^{2}$),
it is almost insensitive to single charged particles and
gamma-ray, but it can be successfully used for EAS particle
density measurements as it will be shown below.

4-channel digital oscilloscope TDS224 connected to a PC via GPIB
interface is used for data acquisition. Integrated analog pulses
from the PMT anode are put to the oscilloscope inputs with
different gain. Digitizing step is equal to 4 $\mu s$ while full
time scale is equal to 10 ms. Full wave form information is
collected in a case of the event.

\section{Experimental results}

The results of this experiment can be found elsewhere
$\cite{st2,st3}$. Here only additional information will be shown
with the aim to illustrate the method performances. First of all,
we have measured the thermal neutron yield per event, which was
found to be equal to $<n>\approx 0.15$ for our event threshold.
The time structure of delayed pulses distribution can be fitted in
an interval of 10 ms by a double-exponential function of a type:
 $F(t)=C(exp(-t/\tau_1)+exp(-t/\tau_2))$ with parameters C, $\tau_1$
 and $\tau_2$ depending on the charged particle density
measured by the ZnS detector and thus depending on the EAS size.
The higher size, the more steep is the time distribution. It is
\begin{figure}[!htb]
\begin{center}
\includegraphics[width=15cm]{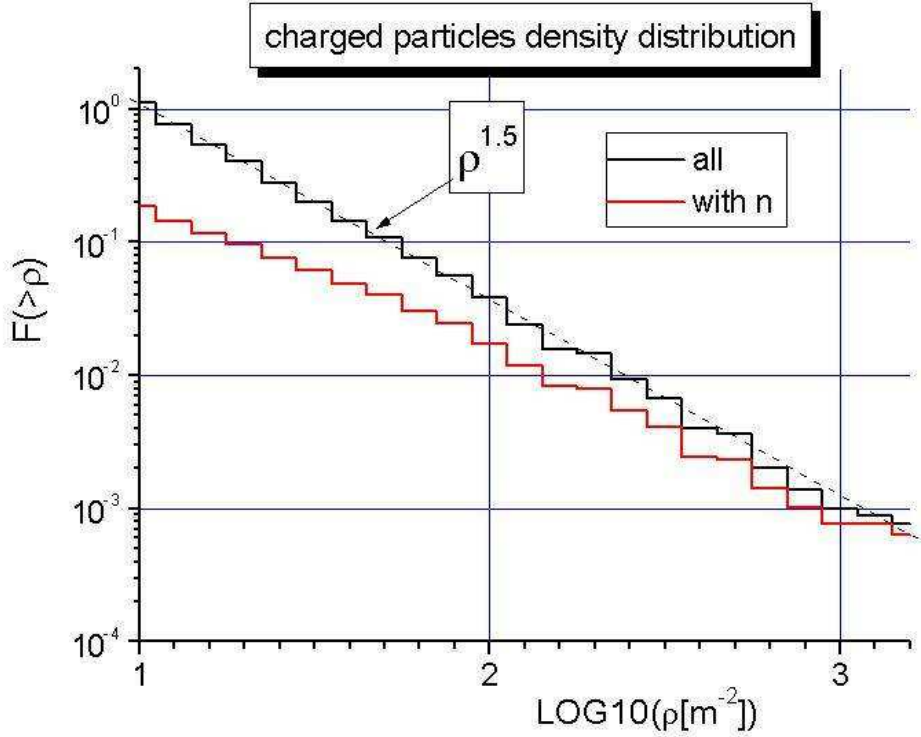}
\caption{Charged particle %
 density distributions measured by the ZnS detector.}
\end{center}
\end{figure}
%
interesting that $\tau_1 \approx 2 \tau_2 \approx 9.0$ ms for all
events and $\approx$ 5.0 ms for events with central density {$\rho
>800 m^{-2}$}. This experimental fact has to be understood and
explained. Unexpectedly flat neutron time distributions could be
explained as follows. As it was mentioned above, the mean
efficiency for neutron detection in our detector is close to 20\%.
But, this value was obtained taking into account spacings between
the scintillator grains, while the efficiency to detect thermal
neutron by the scintillator grain $(0.5 \div 0.8 mm)$ was
calculated to be equal to 74\%. Measured flux is $F=n \times v
\times \varepsilon$, where n is neutron concentration, v is
neutron velocity and $\varepsilon =1-exp(-\lambda t)$ is the
detection efficiency. Here $1/\lambda$ is neutron mean free path
and t is mean grain thickness. It is well known that $\lambda\sim
1/v$ for thermal and slow neutrons. Therefore, $F \approx n$ in a
case of fast neutrons and $F \approx n\times v$ for slow neutrons.
That means that the scintillator is thick enough for thermal
neutrons, while it is thin for fast neutrons. This results in very
flat time distributions (concentration does not changes very
quickly) after the EAS passage and their independence on the media
temperature. In contrast to this case, in a stationary regime when
neutron velocity are in equilibrium with medium (background
measurements) $F\approx n \times v$ and this value depends on the
media temperature  $\cite{ale}$ as $F \sim \sqrt{T}$ due to
Maxwell velocity distribution.
\begin{figure}[!htb]
\hspace{-0.75pc}
\begin{center}
\includegraphics[width=8cm]{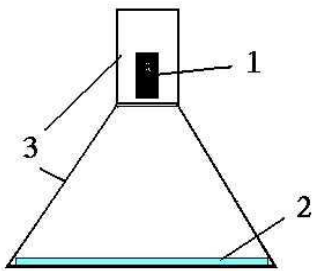}%
\caption{Possible design of the $1-m^{2}$ ZnS detector. %
1 - 6'' PMT ; 2 $-ZnS$+$^{6}Li$ mixture; 3 - housing.}
\end{center}
\end{figure}
A possibility to measure charged particles number passed through
the ZnS detector is demonstrated in the fig. 2, where energy
deposit spectra of the EAS particles are presented as measured by
the ZnS detector . As one can see, the measured particle density
spectrum for all evens follows well known power law function with
integral index equal to $\sim-1.5$, while that for events with
recorded neutrons changes the slope. At high enough density of
about $10^3 ~m^{-2}$ almost all events are with neutrons and the
spectra equalize. These figures confirm that our ZnS detector
works properly not only for neutron detection but for charged
particles as well.

\section{The e-n-array proposal}

The first obtained experimental results as well as results of
Monte Carlo simulations, made on a base of CORSIKA codes (v. 6012,
HDPM and Gheisha models), makes me sure to propose a novel type of
EAS array which could consist only of the large area ZnS detectors
measuring both the main EAS components: hadronic and
electromagnetic. The array could look like a simple grid of say
121 detectors like those shown in fig.3, with spacing $5\div 10$ m
covering an
\begin{figure}[!htb]
\begin{center}
\includegraphics[width=15.cm]{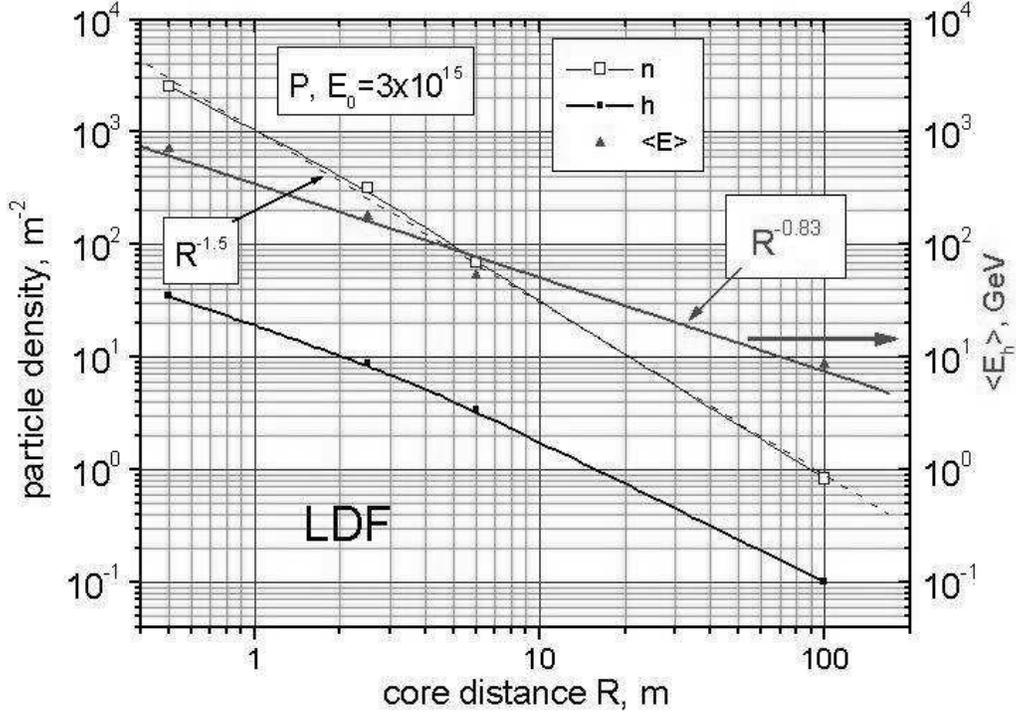}%
\caption{Results of Monte Carlo simulations for hadron and neutron
lateral distribution and mean hadron energy as a function of core
distance for primary proton of 3 PeV.}
\end{center}
\end{figure}
%
area of $100\times100$ $m^{2}$. It could be very informative in
spite of its simplicity and compactness. Detection of thermal
neutron flux accompanying the EAS passage through the surrounding
matter gives absolutely new information, which was never used
before. First of all, the number of detected neutrons is
proportional (in the first approximation) to the number of hadrons
reached the observational level in a radius of $\sim$ 300$\div$500
m around the detector location, including an air layer of the same
thickness. Such a large distance evaporation neutrons can cover
during their movement in air before moderation. Detailed study of
hadronic component with a large area detector ($10^{4}$ $m^{2}$ in
proposed array, which can be extended without any problem) is very
interesting problem because hadrons form the EAS skeleton and only
they can preserve the adequate information about primary particle.
Starting from a low threshold on primary energy of $\sim10\div30$
TeV and covering the ``knee'' region with a wide enough range,
this array could make a significant improvement of experimental
situation and probably would solve the ``knee'' problem. Another
interesting advantage of the array is its possibility to locate
the EAS axis more precisely due to steep lateral distribution of
hadronic and neutron components (see fig.4) in comparison with
electron one usually used for this purpose. That means primary
energy can be recalculated with higher accuracy. A usage of $n/e$
-ratio instead of $\mu/e$-ratio for primary mass composition
measurements would give better results because of first, a number
of thermal neutrons is much higher than a number of muons and
second, electron and hadron components are in equilibrium on an
observational level, while muonic component is not, due to its
integral properties $\cite{st2}$. And finally, the time structure
of the {\it neutron vapor} is absolutely new dimension in EAS
study, which could give us unexpected result.

\section{Summary}

The method to study EAS with thermal neutron scintillator
detectors proposed in 2001 was checked with a one-module
prototype. Promising data obtained with a pioneer experiment let
me to conclude that {\it the neutron vapor} associated with EAS
passage does exist and can provide experimenters with an
additional very useful information. On this basis a novel type of
a very simple array is proposed which could use this additional
dimension in EAS study. A scintillator detector for neutron
detection developed by us showed very good performance and made it
possible to measure thermal neutron flux with very low background.
Moreover, it showed rather good performance in charged particle
density measurements thus lead us to a conclusion that EAS array
could consist only of these detectors measuring both main EAS
components hadronic and electronic. Rather fast pulse width of
$\sim 40$ ns allows one to use these detectors even for the EAS
arrival direction determination if one need not very good angular
resolution. Finally, the detector showed an excellent performance
in thermal neutron background flux measurements. First very
interesting results of this study can be found in $\cite{ale}$.

\section*{Acknowledgements}

The work was supported in part by the RFBR grant N 05-02-17395, by
the Government Contract N 02.445.11.7070  and by the RAS Basic
Research Program ``Neutrino Physics''.

\end{document}